# Диэлектрическая проницаемость при учете вклада электронов в их же экранировку

В.А. Ранцев-Картинов *, А.Н. Кулаков **

*НИЦ Курчатовский институт, E-mail : rankarva@mail.ru

**ФГБНУ Экспертно – Аналитический Центр Министерства Образования и Науки,
E-mail : kulan07@yandex.ru

**Аннотация**

В работе приводится экспериментальный факт наличия ионно-звуковых колебаний в плазме токамака DIII-D при дополнительном нагреве плазмы нейтральным пучком водорода. Это противоречит современным теоретическим представлениям о возможности раскачки таких колебаний в плазме, когда $T_i \geq T_e$, где $T_i$ и $T_e$ - ионная и электронная температура, соответственно. Поскольку электроны обеспечивают свою экранировку, расталкивая вокруг себя электроны же и оголяя ионный остов, то их поляризационное облако должно быть протяженнее дебаевского. Поэтому авторы предложили искать решение кинетического уравнения с учетом вклада электронов в их собственную экранировку, введя некоторый множитель для дебаевского радиуса и, соответственно, разделив на него частоту. В результате получено новое выражение диэлектрической проницаемости, которое переходит в прежнее при равенстве введенного множителя единице. Анализ этого выражения показал, что частотные характеристики возможных низкочастотных и высокочастотных колебаний остались с прежними плазменными частотами, ионно-звуковые колебания получили ту же самую фазовую скорость, но изменилось условие их возбуждения, которое для чистой дейтериевой плазмы приобрело вид: $\frac{T_i}{T_e} \leq 10$, вместо прежнего условия - $\frac{T_e}{T_i} \geq 10$.

# 1. Введение

В условиях эксперимента с дополнительным нагревом нейтральным пучком всегда выполняется условие:

$$T_i \geq T_e, \tag{1}$$

где $T_i$ и $T_e$ - ионная и электронная температура, соответственно. По существующей теории в такой плазме никогда не могут возбуждаться ионно-звуковые колебания, даже в присутствии мощных электронных пучков. Однако, довольно часто в экспериментах с дополнительным нагревом нейтральными пучками при токовой скорости в плазме превышающей ионно-звуковую наблюдается вращение плазмы в направлении протекающего тока со скоростями связанными с этой скоростью. Это может вызвать предположение, что в данной плазме, не смотря на запрет, тем не менее, раскачиваются ионно-звуковые колебания. Отметим здесь, что данный тип неустойчивости имеет очень высокую скорость нарастания амплитуды – порядка величины обратной частоты этих колебаний. Если при этом развилась данного типа турбулентность, то половина электронов плазмы оказывается захваченной этими колебаниями, и они уже движутся не с токовыми скоростями, а с ионно-звуковой скоростью вместе с волной. Эти электроны увлекают за собой соответствующее количество ионов, а чтобы момент сохранился плазма, как целое, начинает вращаться в направлении протекания тока со скоростью ~ $C_s/2$. При этом оказывается, что часть электронов как бы останавливаются, а соответствующая им часть ионов переносит ток с указанной выше токовой скоростью.

Возьмем теперь в качестве примера разряд на американской установке DIII-D № 90117. Омическая мощность в этом разряде составляла ~ $5.3 \cdot 10^5$ Вт, мощность водородного пучка ~ $1.2 \cdot 10^6$ Вт, плотность тока пучка ~ 16 А/см$^2$, энергия частиц пучка ~ 80 кВ, плотность плазмы ~ $4.3 \cdot 10^{13}$ см$^{-3}$, $T_i$ ~ $2.8 \cdot 10^3$ эВ, $T_e$ ~ $2.6 \cdot 10^3$ эВ. Оценка показывает, что плотностью вносимого момента пучком, с точностью до 10% по сравнению с плотностью момента, плазмы вращающейся с половиной звуковой скорости можно пренебречь. Графическое радиальное распределение экспериментально полученной скорости вращения плазмы и расчетной ≈ $C_s/2$ представлено на **Рис. 1.**

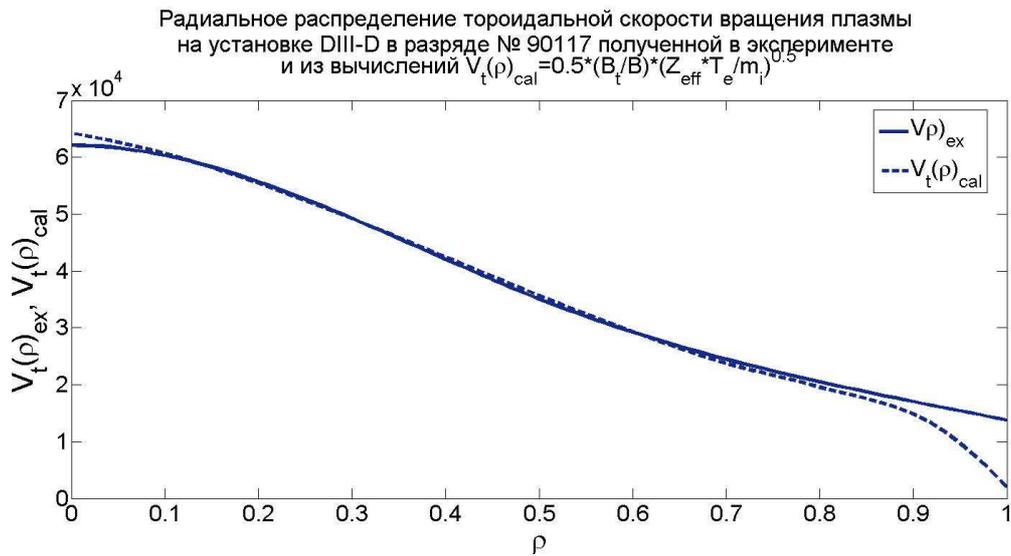

**Рис. 1.** Радиальное распределение тороидальной скорости вращения плазмы на установке DIII-D в разряде № 90117 полученной в эксперименте и из вычислений.

Из анализа этого рисунка видно, что расчетная скорость вращения плазмы (за исключением приосевой и периферийной области) практически совпадает с экспериментально измеренной. В этом разряде токовая скорость превышала ионно-звуковую. Радиальное распределение отношения токовой скорости к ионно-звуковой дано на **Рис. 2**.

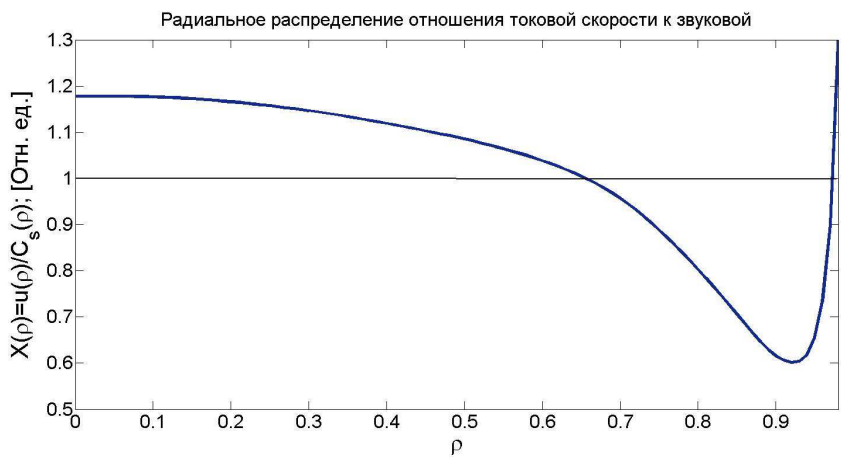

**Рис. 2**. Радиальное распределение токовой скорости к фазовой скорости ионно-звуковых колебаний.

Из анализа этого рисунка видно, что даже в области, где токовая скорость ниже ионно-звуковой, скорость тороидального вращения все равно связана с ней. Очевидно, что к турбулентности причастен также пучок, который имеет скорость, превышающую ионно-звуковую по всему сечению шнура. По классической теории данный случай не мог привести к генерации ионно-звуковой неустойчивости. Для выяснения причины раскачки этих колебаний напомним, какие рассуждения привели к записи выражения диэлектрической проницаемости для двухкомпонентной плазмы.

В общем случае произвольных значений волнового вектора $\vec{k}$, когда существенную роль играет пространственная дисперсия, вычисление проницаемости требует применения кинетического уравнения [1]. Обычно, сначала полагают, что имеют дело только с электронной плазмой, т.е., ионами пренебрегают, считая, что они бесконечно тяжелые и абсолютно неподвижны. В таком случае для слабого поля ищут функцию распределения электронов в виде

$$f^e = f_0^e + \delta f^e, \qquad (2)$$

где $f_0^e$ - невозмущенная полем стационарная изотропная и пространственно-однородная функция распределения электронов, а $\delta f^e$ - ее изменение под влиянием поля. Пренебрегая в кинетическом уравнении членами второго порядка малости, получают:

$$\frac{\partial \delta f^e}{\partial t} + \vec{v}_e \frac{\partial \delta f^e}{\partial \vec{r}} = e\left(\vec{E} + \frac{1}{c}\left[\vec{v}_e \vec{B}\right]\right)\frac{\partial f_0^e}{\partial \vec{p}^e} \qquad (3)$$

Поскольку в изотропной плазме функция распределения зависит только от абсолютной величины импульса, то для такой функции направление вектора $\frac{\partial f_0^e}{\partial \vec{p}^e}$ совпадает с направлением $\vec{p}^e = m_e \vec{v}_e$ и его произведение с $\left[\vec{v}_e \vec{B}\right]$ обращается в нуль. Таким образом, в линейном приближении магнитное поле не влияет на функцию распределения и для изменения функции распределения электронов под влиянием электрического поля, $\delta f^e$, кинетическое уравнение приобретает вид:

$$\frac{\partial \delta f^e}{\partial t} + \vec{v}_e \frac{\partial \delta f^e}{\partial \vec{r}} = e\vec{E}\,\frac{\partial f_0^e}{\partial \vec{p}^e} \qquad (4)$$

## 2. Диэлектрическая проницаемость с частичным учетом динамических эффектов

При выводе диэлектрической проницаемости без учета динамических эффектов, [1], рассматривают отдельно электронную и ионную плазму, находят для них значения диэлектрической проницаемости, а затем их просто складывают, считая, что при этом автоматически учитываются все динамические эффекты, связанные с большой разницей масс электронов и ионов. Однако при таком рассмотрении в экранировке электронов упускается учет вклада самих же электронов плазмы в свою же собственную экранировку. При выводе диэлектрической постоянной электронной плазмы, плазмы с абсолютно неподвижными ионами, считают, что вместе с полем $\vec{E}$ изменение функции распределения электронов, $\delta f^e$, пропорционально экспоненте $\exp[i(\vec{k}\vec{r} - \omega t)]$. Тогда, следуя логике рассуждений проведенной в [1], из (4) получают запись выражения диэлектрической постоянной для электронной составляющей плазмы, находящейся в равновесном состоянии с максвелловским распределением -

$$f_0^e(p_x^e) = \frac{n_e}{(2\pi n_e T_e)^{1/2}} \exp\left(-\frac{(p_x^e)^2}{2m_e T_e}\right), \qquad (5)$$

где $T_e$ температура электронов плазмы:

$$\varepsilon_l(\omega, k) = 1 + \frac{1}{(k a_e)^2}\left[\text{Re}\left(1 + F\left(\frac{\omega}{\sqrt{2} k v_{T_e}}\right)\right) + \text{Im}\left(F\left(\frac{\omega}{\sqrt{2} k v_{T_e}}\right)\right)\right]. \qquad (6)$$

Здесь

$$v_{T_e} = \sqrt{\frac{T_e}{m_e}}, \; a_e = \sqrt{\frac{T_e}{4\pi n_e e^2}}, \qquad (7)$$

$v_{T_e}$ - некоторая средняя тепловая скорость электронов, $a_e$ - дебаевский радиус статического экранирования электронов. Если ввести обозначение -

$$x = \frac{\omega}{\sqrt{2} k v_{T_e}}, \qquad (8)$$

тогда функция $F(x)$ определена интегралом

$$F(x) = \frac{x}{\sqrt{\pi}} \int_{-\infty}^{\infty} \frac{e^{-z^2} dz}{z - x - i0} = \frac{x}{\sqrt{\pi}} \int_{-\infty}^{\infty} \frac{e^{-z^2} dz}{z - x} + i\sqrt{\pi} x e^{-x^2}, \qquad (9)$$

которая может быть выражена через интеграл вероятностей $W(x)$ -

$$F(x) = i\sqrt{\pi} W(x), \qquad (10)$$

который приведен в виде таблиц для различных значений аргумента и его мнимой составляющей в справочнике [2]. Предельные значения функции $F(x)$ для больших и малых значений $x$ даны ниже:

$$F(x) + 1 \approx -\frac{1}{2x^2} + i\sqrt{\pi} x e^{-x^2}, \text{ при } x \gg 1;$$

$$\text{и } F(x) \approx -2x^2 + i\sqrt{\pi} x, \text{ при } x \ll 1. \qquad (11)$$

Графические представления функции $Z(x) = \mathrm{Re}(F(x) + 1)$, определяющей диэлектрическую проницаемость (6), и функции $z(x) = 0.5 * \mathrm{Im}(F(x))$, фактически определяющей декремент затухания продольных колебаний плазмы, $\gamma$ (см. (12)), приведены на **Рис. 1**.

$$\gamma = 0.5 \cdot \omega \cdot \mathrm{Im}(\varepsilon_l(\omega, k)) = 0.5 \cdot \omega \cdot \frac{1}{(k a_e)^2} \mathrm{Im}(F(x)) = \frac{\omega}{(k a_e)^2} z(x), \qquad (12)$$

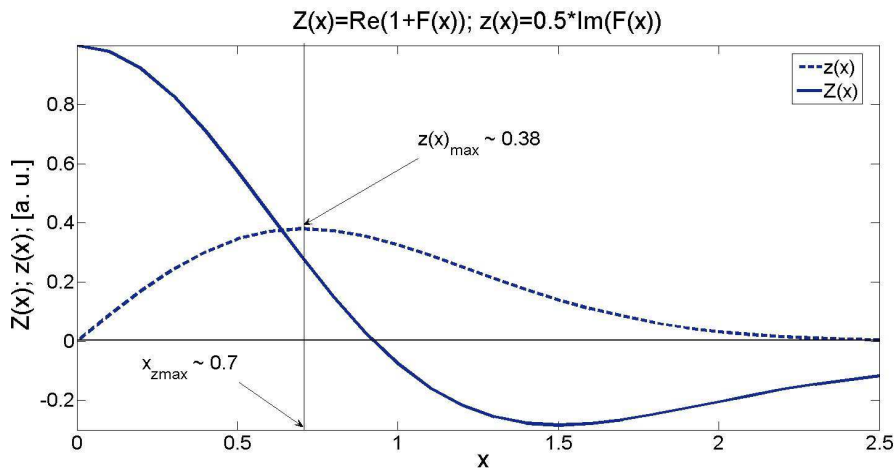

**Рис. 3**. Даны графические представления функции $Z(x) = \text{Re}(1 + F(x))$, входящей в определение продольной диэлектрической проницаемости (6), а также функции $z(x) = 0.5 \cdot \text{Im}(F(x))$, определяющей декремент затухания продольных колебаний (12).

Поскольку функция $z(x) \leq 0.38$ для любых значений ее аргумента, то в плазме, в той или иной мере, могут присутствовать коротковолновые колебания, для которых выполняется условие

$$ka_e \geq 1, \text{ т.е., } \lambda \leq a_e. \tag{13}$$

Однако, такие колебания быстро затухают вне дебаевской сферы, а потому они не распространяются в плазме и далее нами из рассмотрения исключаются.

Вклад ионной составляющей в диэлектрическую проницаемость вычисляется точно таким же способом, полагая неподвижными электроны, и затем оба вклада в виде $\varepsilon_l - 1$ просто складываются. Таким образом, формула для диэлектрической проницаемости плазмы с учетом электронного и ионного вкладов при таком рассмотрении записывается как:

$$\varepsilon_l(\omega, k) = 1 + \frac{1}{(ka_e)^2}\left[1 + F\left(\frac{\omega}{\sqrt{2}kv_{T_e}}\right)\right] + \frac{1}{(ka_i)^2}\left[1 + F\left(\frac{\omega}{\sqrt{2}kv_{T_i}}\right)\right], \tag{14}$$

где $v_{Ti}$ - средняя тепловая скорость ионов и $a_i$ - дебаевский радиус статического экранирования ионов, которые определены выражениями:

$$v_{T_i} = \sqrt{\frac{T_i}{m_i}}, \ a_i = \sqrt{\frac{T_i}{4\pi Z_i n_e e^2}}, \tag{15}$$

а $Z_i$ - заряд ионов. Выражение (14) относится к плазме двумя разными температурами – электронной и ионной. Каждая из компонент этой плазмы имеет равновесное распределение, но с различными температурами, поскольку электроны с ионами не могут установить равновесного состояния из-за большой разницы в их массах, затрудняющей обмен энергией между ними при столкновениях. Данное выражение уже частично учитывает динамические эффекты, но не полностью, поскольку при таком рассмотрении не учитываются вклады одноименных с рассматриваемыми кернами поляризационных облаков зарядов плазмы. Если для экранировки ионов таким вкладом почти всегда можно пренебречь, ввиду их малой подвижности, то для электронов этого делать не следует, ибо этот вклад может оказаться существенным. Итак, в выражении (14) во втором его члене не учтен вклад электронов в их же экранировку. Это упущение мы попытаемся подправить в следующем параграфе, и посмотреть к чему это приведет.

### **3. Более полный учет динамических эффектов**

Как следует из работ [3-6], при учете динамических эффектов, связанных с учетом вклада электронов в экранировку электронов, пространственный и временной масштабы значительно увеличиваются. Это происходит из-за большой разницы в массах положительных ионов плазмы и электронов. Если рассматривать движение ионов плазмы с реальной разницей их температуры по сравнению с температурой электронной компоненты, то электроны, обладая очень высокой подвижностью по сравнению с ионами, всегда успевают отследить их координату и перераспределиться так, чтобы обеспечить каждому иону плазмы дебаевскую экранировку. Иная ситуация возникает, когда мы пытаемся рассмотреть процесс экранирования движущихся электронов плазмы. Из-за большой разницы в подвижности, ионы не успевают перераспределяться так быстро, чтобы обеспечить экранировку движущемуся электрону плазмы в каждой точке его траектории. Поэтому рассматриваемый нами движущийся электрон, чтобы обеспечить себя экранировкой, расталкивает вокруг себя электроны плазмы в виде некоторой сферической оболочки, и на оголившемся распределении ионов внутри этой оболочки и на электронах тела самой оболочки строит экранирующее поляризационное облако. Очевидно, полный размер такого поляризационного облака динамической экранировки электрона окажется больше дебаевского.

На **Рис.4** представлена относительная величина модуля усредненного эффективного заряда, $|<Q(x)>|$, движущегося в плазме с тепловой скоростью электрона $v$ ($\mu \equiv v/v_{T_e} \approx 1$, здесь $v_{T_e}$ - средняя тепловая скорость электронов), как функция расстояния от этого заряда выраженная в единицах радиуса Дебая [5] $\left( x = r/r_{D_e} \right)$, расчетная формула, которая дана в работе [4].

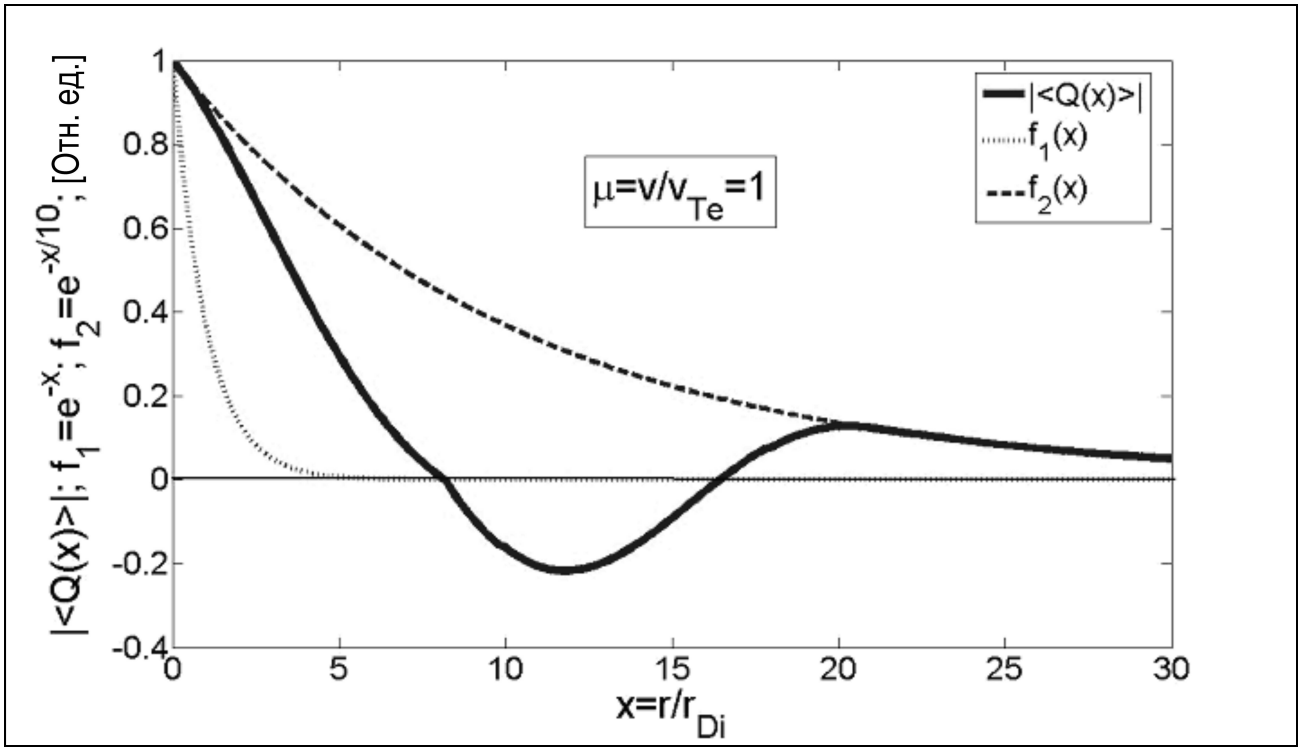

**Рис. 4.** Представлена относительная величина усредненного эффективного заряда $\langle Q(r) \rangle$ (сплошная кривая) движущегося в плазме, в данном случае, электрона, для значений $\mu \equiv v/v_{T_e} \approx 1$, как функция расстояния от этого заряда выраженная в единицах соответствующего дебаевского радиуса, (в данном случае $r_{D_i}$, и $x = r/r_{D_i}$) [4]. $v$ - здесь конкретная скорость рассматриваемого электрона, а $v_{T_e}$ соответствует его тепловой скорости в плазме. Точечная кривая здесь соответствует дебаевской экранировке электрона, т.е., кривой $f_1 = e^{-x}$. Пунктирная кривая на рисунке соответствует экранировке заряда с радиусом поляризационного облака, который в 10 раз больше дебаевского, т.е., $f_2 = e^{-x/10}$.

Чтобы избежать решения сложной нелинейной задачи, попробуем учесть феноменологическое увеличение размера поляризационного облака электрона путем введения некоторого коэффициента $\zeta$. Далее, при рассмотрении динамических эффектов электронной составляющей плазмы положим, что вместе с полем $\vec{E}$ функция $\delta f^i$ пропорциональна $\exp[i\zeta^2(\vec{k}\vec{r} - \zeta^{-1}\omega)]$, в то время как в общем случае вместе с полем $\vec{E}$ изменение функции распределения ионов $\delta f^i$ будет пропорциональна $\exp[i(\vec{k}\vec{r} - \omega)]$, т.е., в этом случае для ионной компоненты плазмы мы имеем:

$$\delta f^i = \frac{e\vec{E}}{i\zeta^2\left(\vec{k}\vec{r} - \dfrac{\omega}{\zeta}\right)} \frac{\partial f_0^i}{\partial \vec{p}^i} \qquad (16)$$

Плотность зарядов электронов в каждой точке компенсируется зарядами ионов, и при ее возмущении электрическим полем также описывается соответствующим выражением из [1]. Как и $\delta f^i$ величина этого поля будет пропорциональна $\exp\left[i\zeta^2\left(\vec{k}\vec{r} - \dfrac{\omega}{\zeta}\right)\right]$, а $\zeta$ - введенный нами параметр, который нам еще предстоит определить.

Далее, следуя рассуждениям [1], приведенным при выводе формулы (6), получим выражение вклада ионов в диэлектрическую проницаемость при учете динамических эффектов, связанных с экранированием электронов –

$$\varepsilon_l(\omega,k) = 1 + \frac{1}{(\zeta^2 k a_i)^2}\left[1 + F\left(\frac{\omega}{\sqrt{2}\zeta k v_{T_e}}\right)\right], \qquad (17)$$

где функция $F(x)$ определена выражением (9), а ее предельные значения даны в (11).

Поскольку при любом значении температуры электроны настолько подвижны по сравнению с ионами, что они успевают отследить ионы вдоль их траектории так, чтобы обеспечить им дебаевское экранирование, то поправка в экранировку ионов при учете нами динамических эффектов незначительна, и вклад электронов в экранировку ионов остается прежним. Поэтому, отличие от (14) в обобщенном выражении диэлектрической постоянной при учете электронной и ионной компоненты с учетом динамических эффектов касается только второго члена данного выражения, представленного ниже:

$$\varepsilon_l(\omega,k) = 1 + \frac{1}{(k a_e)^2}\left[1 + F\left(\frac{\omega}{\sqrt{2} k v_{T_e}}\right)\right] + \frac{1}{(\zeta^2 k a_i)^2}\left[1 + F\left(\frac{\omega}{\sqrt{2}\zeta k v_{T_i}}\right)\right]. \qquad (18)$$

## 4. Низкочастотные колебания

Рассмотрим сначала значение (18) при низких частотах, т.е., когда выполняется условие:

$$\frac{\omega}{k} \ll \sqrt{2}\zeta v_{Ti} \ll \sqrt{2} v_{Te}. \tag{19}$$

Из (18) имеем:

$$\varepsilon_l(\omega,k) = 1 + \frac{1}{(ka_e)^2}\left[1-\left(\frac{\omega}{kv_{Te}}\right)^2 + i\sqrt{\frac{\pi}{2}}\frac{\omega}{kv_{Te}}\right] + \frac{1}{(\zeta ka_i)^2}\left[1-\left(\frac{\omega}{\zeta kv_{Ti}}\right)^2 + i\sqrt{\frac{\pi}{2}}\frac{\omega}{\zeta kv_{Ti}}\right] \approx$$

$$\approx 1 + \frac{1}{(ka_e)^2} + \frac{1}{(\zeta ka_i)^2} + i\sqrt{\frac{\pi}{2}}\frac{\omega}{k^3 a_e^3 \Omega_e} + i\sqrt{\frac{\pi}{2}}\frac{\omega}{\zeta^3 k^3 a_i^3 \Omega_i}. \tag{20}$$

Чтобы определить величину параметра $\zeta$, рассмотрим колебания с предельными частотами $\omega \to 0$. Такие колебания могут иметь место только тогда, когда мнимые части выражения (20) равны. В этом случае колебания электронов и ионов в поляризационном облаке происходит с одной и той же частотой и из выполнения этого условия определяется значение $\zeta$.

$$\zeta = \frac{a_e}{a_i}\left(\frac{\Omega_e}{\Omega_i}\right)^{1/3} = Z_{eff}^{1/3}\left(\frac{T_e}{T_i}\right)^{1/2}\left(\frac{m_i}{m_e}\right)^{1/6} \tag{21}$$

Для чистой дейтериевой плазмы, если $T_e \sim T_i$ из (21) получаем:

$$\zeta_D = \left(\frac{m_D}{m_e}\right)^{1/6} \approx 3.93 \tag{22}$$

Общий вид значения коэффициента $\zeta$, учитывающего случай статического и динамического рассмотрения экранировки ионов в чистой дейтериевой при равенстве температур электронов и ионов плазмы можно записать следующим образом:

$$\zeta^2 = \left(\left(\frac{m_D}{m_e}\right)^{1/3} - 1\right)\left(\frac{m_D}{m_i}\right)^{1/3} + 1 = \left(\zeta_D^2 - 1\right)\left(\frac{m_D}{m_i}\right)^{1/3} + 1. \qquad (23)$$

При статическом рассмотрении в (23) $m_i \to \infty$ и $\zeta \to 1$, при учете же динамических эффектов в этом выражении нужно положить $m_i \equiv m_D$ и тогда $\zeta \equiv \zeta_D$.

## 5. Потенциал покоящегося пробного заряда в равновесной плазме с учетом динамической поправки

При условии нулевых частот диэлектрическая проницаемость (18) (как и в случае без учета динамических эффектов) описывает экранирование малого пробного точечного покоящегося заряда $q$ в плазме. Действительно, с учетом поляризации плазмы, электрическое поле определяется уравнением:

$$div\vec{D} = 4\pi q \delta(\vec{r}). \qquad (24)$$

Для нулевых частот Фурье-компоненты индукции и потенциала связаны соотношением

$$\vec{D}_{\vec{k}} = \varepsilon_l(0,k)\vec{E}_{\vec{k}} = -i\vec{k}\varepsilon_l(0,k)\varphi_{\vec{k}} \qquad (25)$$

Из (25) получаем уравнение для определения $\varphi_{\vec{k}}$, которое с учетом (21 и 22) записывается в виде

$$i\vec{k}\vec{D}_{\vec{k}} = k^2\varepsilon_l(0,k)\varphi_{\vec{k}} = \left[k^2 + a_e^{-2}\left(1 + \frac{Z_{eff}T_e}{\zeta^2 T_i}\right)\right] = \left[k^2 + a_e^{-2}\left(1 + \frac{Z_{eff}^{1/3}}{15.44}\right)\right] = 4\pi q \qquad (26)$$

Теперь, взяв $\varepsilon_l(0,k)$ из (26), получаем

$$\varphi_{\vec{k}} = \frac{4\pi q}{\left[ k^2 + a_e^{-2}\left(1 + \frac{Z_{eff}^{1/3}}{15.44}\right)\right]} \qquad (27)$$

Мы видим, что уже здесь выпадает зависимость потенциала от соотношения электронной и ионной температуры. Т.е., учтенная нами динамическая поправка делает экранировку покоящихся пробных частиц в двух температурной плазме (через дебаевский радиус) зависимой только от электронной температуры. Обратное Фурье-преобразование (27) дает координатную функцию распределения потенциала –

$$\varphi = \frac{q}{r}\exp(-\frac{r}{a}) = \frac{q}{r}\exp\left[-\frac{r}{a_e\left(1 + \frac{Z_{eff}^{1/3}}{15.44}\right)^{-1/2}}\right]. \qquad (28)$$

Здесь малость точечного заряда $q$ предполагает, что его величина должна быть мала по сравнению с зарядом частиц плазмы в объеме $\sim a_e^3$. Таким образом, мы видим, что дополнительный учет динамических эффектов меняет распределение потенциала электрического поля вокруг заряда, даже если он покоится, ибо в этом случае, согласно (28),

$$a = a_e\left(1 + \frac{Z_{eff}^{1/3}}{15.44}\right)^{-1/2} \sim a_e \qquad (29)$$

Согласно (29), если мы имеем дело с дейтериевой чистой плазмой, радиус экранирования неподвижного заряда не зависит от соотношения электронной и ионной температуры плазмы и всегда почти равен $a_e$. Здесь следует заметить, что при рассмотрении статической задачи, электронной плазмы, $a \sim a_e$, а в ионной плазме $a \sim a_i$. Иными словами, выражение (14) уже, до некоторой степени, учитывает динамические эффекты, но не полностью, поскольку не учитывает вклад электронов в экранировку самих же электронов. Рассмотрение случая низкочастотных колебаний для (14) приводит нас к выражению радиуса экранирования

$$a = a_e \left(1 + \frac{Z_{eff} T_e}{T_i}\right)^{-1/2}, \qquad (30)$$

которое при $\frac{Z_{eff} T_e}{T_i} \gg 1$, дает $a \sim a_i$, что его сильно отличает это выражение от (29).

## 6. Потенциал движущегося пробного заряда в равновесной плазме с учетом динамической поправки

Совсем другая картина наблюдается, когда рассматривается потенциал поля движущегося пробного заряда, ибо в этом случае эффект экранирования уменьшается [3-6]. Кроме того, при движении заряда появляется анизотропия углового распределения его потенциала, которая исчезает как при рассмотрении малых (по сравнению с тепловой скоростью электронов плазмы), так и больших скоростей заряда. Здесь предполагается, что рассматриваемые скорости далеки от скорости света. В работах [7, 8] было показано, что при движении системы, состоящей из стороннего заряда и создаваемого им поляризационного облака, у нее появляется квадрупольный момент, который становится равным нулю только при нулевой скорости стороннего заряда, так что вне поляризационного облака потенциал системы убывает пропорционально кубу расстояния. Исследованию особенностей проявления динамического экранирования, как собственных частиц равновесной плазмы, так и движущихся в ней сторонних зарядов посвящено много работ: общие выражения для стационарного электрического поля, движущегося с постоянной скоростью в изотропной плазме точечного заряда, даны в [9-11]; численные расчеты с применением модельной дисперсионной функции были проведены авторами [12, 13]; детальные исследования проявлений динамического экранирования представлены в работах [3, 4, 14, 15].

Следуя подходу, изложенному в работах [4, 16-19], будем использовать потенциал частицы с зарядом $e_a$ движущейся с постоянной скоростью в плазме, $\vec{v}$ (предполагается движение нерелятивистским). В частности, таким пробным зарядом можно считать любую частицу самой плазмы. Выражение этого потенциала имеет вид, [9]:

$$\varphi_{a\vec{r}(t)\vec{v}}(\vec{r}) = \frac{e_a}{2\pi^2} \int \frac{d\vec{k}}{k^2} e^{i\vec{k}(\vec{r}-\vec{r}(t))} \int_{-\infty}^{\infty} \frac{d\omega \delta(\omega - \vec{k}\vec{v})}{\varepsilon_l(\vec{k},\omega)}, \qquad (31)$$

здесь

$$\vec{r}(t) = \vec{r}_0 + \vec{v}t \qquad (32)$$

Следует отметить, что характер экранирования плазмой движущегося заряда качественно отличается от дебаевского экранирования неподвижного заряда, что проявляется в не экспоненциальном убывании потенциала (31) с расстоянием $\vec{r}(t)$. Это можно обнаружить при рассмотрении медленных движений, когда $v \ll v_a$, определяемой выражением (31). Можно показать, что после разложения (31) с точностью до линейного члена по скорости заряда $v$ и, учитывая условие низких частот, $\frac{\omega}{\sqrt{2}kv} \ll 1$ в выражении диэлектрической проницаемости с учетом динамических эффектов (20), на расстояниях $r \gg a_e$ значение потенциала приобретает вид:

$$\varphi(r) = q\left\{\frac{1}{r}\exp\left[-\frac{r}{a_e\left(1+\frac{Z_{eff}^{1/3}}{15.44}\right)^{-1/2}}\right] + 4(2\pi)^{1/2}\frac{a_e^2(\vec{v}\vec{r})}{\left(1+\frac{Z_{eff}^{1/3}}{15.44}\right)v_e r^3}\right\}. \qquad (33)$$

Графики нормированных на единицу при $x=10$, где $x=r/a_e$, соответствующих функциональных зависимостей обобщенных зарядов даны на **Рис. 5**.

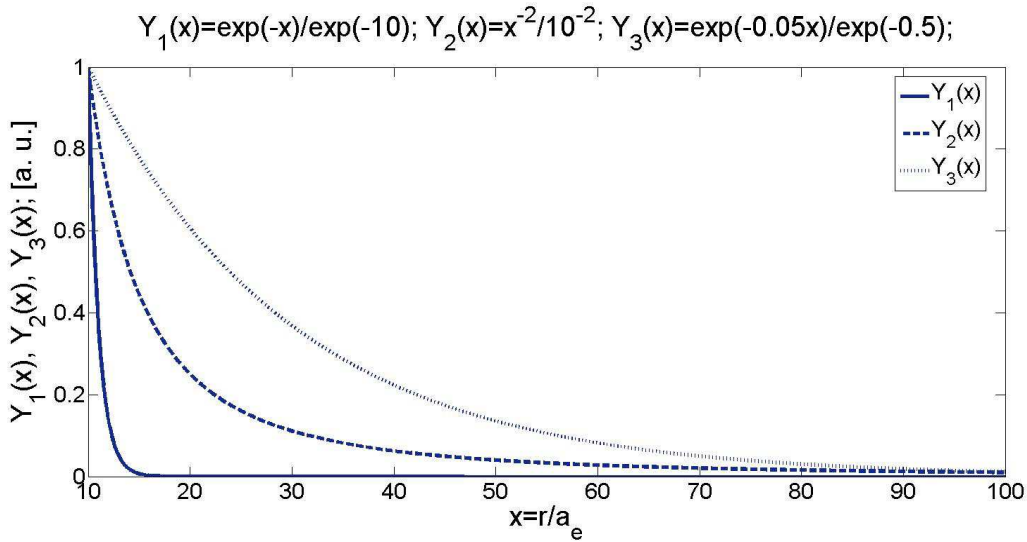

**Рис. 5**. Представлены графики нормированных на единицу множителей при кулоновском потенциале при $x = 10$, где $x = r/a_e$, экспоненциальной части потенциала (33), полученной при $v = 0.1 v_e$,

$$Y_1(x) = \frac{\exp\left[-\dfrac{x}{\left(1+\dfrac{Z_{eff}^{1/3}}{15.44}\right)^{-1/2}}\right]}{\exp\left[-\dfrac{10}{\left(1+\dfrac{Z_{eff}^{1/3}}{15.44}\right)^{-1/2}}\right]} \approx \frac{\exp(-x)}{\exp(-10)}$$

и соответствующей правой части (33),

$Y_2(x) = \dfrac{x^{-2}}{10^{-2}}$, а также соответствующего экспоненциального множителя $Y_3(x) = \dfrac{\exp(-0.05x)}{\exp(-0.5)}$, на очень больших расстояниях совпадающего с $Y_2(x)$ и имеющего показатель экспоненты соответствующего радиусу экранирования $\sim 20 \cdot a_e$.

Из анализа **Рис. 5** видно, что на расстояниях $r \gg 10 \cdot a_e$ спад функции обобщенного заряда $Y_3$, отвечающего потенциалу (33), полученного при $v = 0.1 \cdot v_e$, соответствует показателю экспоненты $\sim 20 \cdot a_e$, т.е., учет движения пробной частицы даже с небольшой скоростью приводит к существенной разъэкранировке дебаевского экранирования на больших расстояниях. Таким образом, на больших (по сравнению с радиусом Дебая) расстояниях потенциал движущегося заряда в плазме спадает обратно пропорционально кубу расстояния и, кроме того, существенно зависит от направления. При этом на больших расстояниях первым слагаемым, т.е., кулоновской составляющей потенциала, по сравнению со вторым, можно пренебречь, а экранировка на расстояниях $r \gg 10 \cdot a_e$ затягивается на расстояния в десятки раз превышающие электронный дебаевский радиус.

### 7. Высокочастотные колебания

Высокочастотные колебания определяются выполнением условия:

$$\frac{\omega}{k} \gg \sqrt{2} v_{T_e} \geq \sqrt{2} \zeta v_{T_i}. \tag{34}$$

В этом случае (18) переписывается в виде:

$$\varepsilon_l(\omega,k) = 1 + \frac{1}{(ka_e)^2}\left[-\left(\frac{kv_{Te}}{\omega}\right)^2 - \frac{3}{4}\left(\frac{kv_{Te}}{\omega}\right)^4 + i\sqrt{\frac{\pi}{2}}\frac{\omega}{kv_{Te}}e^{-\left(\frac{\omega}{2kv_{Te}}\right)^2}\right] +$$

$$+ \frac{1}{(\zeta ka_i)^2}\left[-\left(\frac{\zeta kv_{Ti}}{\omega}\right)^2 - \frac{3}{4}\left(\frac{\zeta kv_{Ti}}{\omega}\right)^4 + i\sqrt{\frac{\pi}{2}}\frac{\omega}{\zeta kv_{Ti}}e^{-\left(\frac{\omega}{\sqrt{2}\zeta kv_{Ti}}\right)^2}\right] \approx$$

$$\approx 1 - \left(\frac{\Omega_e}{\omega}\right)^2\left[1 + 3\left(\frac{kv_{Te}}{\omega}\right)^2\right] + i\sqrt{\frac{\pi}{2}}\frac{\omega\Omega_e^2}{(kv_{Te})^3}e^{-\left(\frac{\omega}{\sqrt{2}kv_{Te}}\right)^2} -$$

$$- \left(\frac{\Omega_i}{\omega}\right)^2\left[1 + 3\left(\frac{\zeta kv_{Ti}}{\omega}\right)^2\right] + i\sqrt{\frac{\pi}{2}}\frac{\omega\Omega_i^2}{(\zeta kv_{Ti})^3}e^{-\left(\frac{\omega}{\sqrt{2}\zeta kv_{Ti}}\right)^2}. \tag{35}$$

Здесь

$$\Omega_e = \frac{v_{Te}}{a_e} = \left(\frac{4\pi n_e e^2}{m_e}\right) \tag{36}$$

Если $3\left(\frac{kv_{Te}}{\omega}\right)^2$, $3\left(\frac{\zeta kv_{Ti}}{\omega}\right)^2 \ll 1$, тогда (35) можно переписать в виде:

$$\varepsilon_l(\omega,k) \approx 1 - \left(\frac{\Omega_e}{\omega}\right)^2 - \left(\frac{\Omega_i}{\omega}\right)^2 + i\sqrt{\frac{\pi}{2}}\frac{\omega\Omega_e^2}{(kv_{Te})^3}e^{-\left(\frac{\omega}{\sqrt{2}kv_{Te}}\right)^2} \approx 1 - \left(\frac{\Omega_e}{\omega}\right)^2 + i\sqrt{\frac{\pi}{2}}\frac{\omega\Omega_e^2}{(kv_{Te})^3}e^{-\left(\frac{\omega}{\sqrt{2}kv_{Te}}\right)^2}. \tag{37}$$

Частота продольных колебаний определяется, при выполнении условия

$$\varepsilon_l(\omega,k) \equiv 0 \tag{38}$$

и при вышеуказанных условиях (34) она просто равна электростатической плазменной частоте и, в первом приближении, не зависящей от $k$.

$$\omega \equiv \Omega_e. \qquad (39)$$

Этот тип колебаний являются длинноволновыми, поскольку имеют место только при выполнении условия:

$$\frac{\omega}{\sqrt{2}kv_{Te}} = \frac{\Omega_e \lambda}{\sqrt{2}v_{Te}} = \frac{\lambda}{\sqrt{2}a_e} \gg 1, \qquad (40)$$

что определяет допустимые длины волн таких колебаний

$$\lambda \gg \sqrt{2}a_e \qquad (41)$$

Т.е., длинноволновые высокочастотные плазменные колебания имеют длину волн $\lambda \sim 14 \cdot a_e$, а потому они могут вызывать когерентное смещение заряженных частиц поляризационных облаков относительно их кернов, вызывая тем самым (при наличии градиентов параметров плазмы) появление локального электрического поля, формирующегося на микро масштабе, а проявляющегося в макроскопических полях. В первом приближении декремент затухания этих плазменных волн определяется выражением:

$$\gamma = \sqrt{\frac{\pi}{8}} \frac{\Omega_e}{(ka_e)^3} \exp\left(-\frac{1}{2(ka_e)^2}\right). \qquad (42)$$

В силу (40)) он действительно оказывается экспоненциально малым и возрастает с уменьшением длины волны. При $ka_e \sim 1$ (когда формула (42) уже неприменима) он становится порядка частоты и понятие о распространяющихся плазменных волнах теряет всякий смысл.

## 8. Промежуточные частоты

Рассмотрим теперь случай промежуточных частот, а, именно, когда выполняется неравенство:

$$\sqrt{2}\zeta v_{Ti} \ll \frac{\omega}{k} \ll \sqrt{2} v_{Te}. \qquad (43)$$

В этом случае (18) переписывается в виде:

$$\varepsilon_l(\omega,k) = 1 + \frac{1}{(ka_e)^2}\left[1 - \left(\frac{\omega}{kv_{Te}}\right)^2 + i\sqrt{\frac{\pi}{2}}\frac{\omega}{kv_{Te}}\right] + \frac{1}{(\zeta^2 ka_i)^2}\left[-\left(\frac{\zeta kv_{Ti}}{\omega}\right)^2 + i\sqrt{\frac{\pi}{2}}\frac{\omega}{\zeta kv_{Ti}}e^{-\left(\frac{\omega}{\sqrt{2}\zeta kv_{Ti}}\right)^2}\right] \approx$$

$$\approx 1 + \frac{1}{(ka_e)^2} - \left(\frac{\Omega_i}{\omega}\right)^2 + i\sqrt{\frac{\pi}{2}}\frac{\omega}{(\zeta ka_i)^3 \Omega_i}e^{-\left(\frac{\omega}{\sqrt{2}\zeta kv_{Ti}}\right)^2}. \qquad (44)$$

Здесь введено обозначение

$$\Omega_i = \frac{v_{Ti}}{a_i} = \left(\frac{4\pi Z n_e e^2}{m_i}\right). \qquad (45)$$

В случае предельно больших длин волн,

$$\frac{1}{(ka_e)^2} \gg 1, \qquad (46)$$

что соответствует длинам, волн удовлетворяющим условию

$$\lambda = \frac{1}{k} \geq 10^{1/2}\cdot a_e \sim 3.16\cdot a_e. \qquad (47)$$

Частота возможных продольных колебаний определяется из выполнения условия (47), которая оказывается пропорциональной волновому вектору $k$, что характерно для звуковых колебаний. Из (44) находится значение фазовой скорости этих колебаний

$$\frac{\omega}{k} = \sqrt{Z_{eff}\frac{T_e}{m_i}}. \tag{48}$$

Из этого выражения видно, что найденная скорость равна скорости распространения ионного звука, полученного при классическом рассмотрении, т.е., когда выполняется условие –

$$\zeta \equiv 1. \tag{49}$$

Оценка мнимой части выражения (44) при учете (43) дает условие существования ионно-звуковых колебаний с фазовой скоростью (48):

$$\sqrt{\frac{\pi}{8}}\frac{\omega}{(\zeta k a_i)^3 \Omega_i} e^{-\left(\frac{\omega}{\sqrt{2}\zeta k v_{Ti}}\right)^2} = \sqrt{\frac{\pi Z_{eff} T_e}{8 T_i}} \frac{1}{3.93^3 Z_{eff}} \left(\frac{T_i}{T_e}\right)^{3/2} \left(\frac{\lambda}{a_e}\right)^2 e^{-\frac{Z_{eff}}{24.5_i}} \approx$$

$$\approx \sqrt{\frac{\pi}{8 Z_{eff}}} \frac{T_i}{T_e} \frac{1}{3.93^3} \left(\frac{\lambda}{a_e}\right)^2 \le 1, \tag{50}$$

что с учетом (47) приводит к неравенству

$$\sqrt{\frac{\pi}{8 Z_{eff}}} \frac{T_i}{T_e} \frac{1}{3.93^3} \le 0.1, \tag{51}$$

Отсюда получаем ограничение на соотношение температур электронов и ионов в чистой дейтериевой плазме.

$$\frac{T_i}{T_e} \le 0.1 \cdot 3.93^3 \sqrt{\frac{8}{\pi}} \sim 9.7 \sim 10 \tag{52}$$

Для исследуемых нами ионно-звуковых колебаний в чистой дейтериевой плазме длину их волн определяет выражение (47), а частота, соответственно, выражением (53):

$$\omega \le \frac{\Omega_i}{10^{1/2}} \sim \frac{\Omega_i}{3.16}. \tag{53}$$

С другой стороны, в области малых длин волн, когда выполняется условие:

$$Z\frac{T_e}{T_i} \gg 1, \quad (54)$$

в (44) можно пренебречь вторым членом справа и тогда, учитывая, что для колебаний необходимо выполнение (38), имеем:

$$\omega = \Omega_i. \quad (55)$$

Это и есть ионные плазменные колебания. Ограничение на их длины волн можно выразить неравенством:

$$\lambda \geq 10 \cdot a_i. \quad (56)$$

## 9. Заключение

Итак, эффекты, связанные с учетом участия электронов в экранировке электронов же плазмы не вносит каких-либо существенных изменений в результаты расчетов колебаний. Действительно, в рассматриваемом случае высокочастотные и низкочастотные колебания раскачиваются с теми же, что и прежде, соответствующими плазменными частотами, и значение фазовой скорости ионно-звуковых колебаний также осталась неизменным. Изменение получило лишь только условие существования ионно-звуковых колебаний. *При раскачке ионно-звуковых колебаний, учет рассматриваемой в работе динамической поправки, снимает условие необходимости выполнения условия*

$$T_e \gg T_i. \quad (57)$$

Но только это обстоятельство уже может существенно изменить сценарий развития физических процессов в плазме, поскольку, если есть механизм раскачки таких колебаний (например, наличие пучков или токовой скорости, превышающей ионно-звуковую скорость), то они всегда будут иметь место, так как в экспериментальной плазме условие (52) всегда выполнено. При этом их фазовая скорость определяется выражением (48), допустимые длины волн выражением (47) и допустимые частоты – выражением (53). Эти колебания, как и высокочастотные, являются длинноволновыми. Потому они также могут вызывать когерентное смещение поляризационных облаков заряженных частиц относительно экранируемых ими кернов, вызывая (при наличии градиентов параметров плазмы) появление макроскопических электрических полей, механизм зарождения которых является микроскопическим, а их проявление носит макроскопический характер.

Таким образом, полный учет динамических эффектов вносит изменения только в условия существования колебаний для случая промежуточных частот определяемых выражением (43). Действительно, предельный переход в выражении (23) при $m_i \to \infty$ соответствует прежнему рассмотрению задачи, когда $\zeta \to 1$, т.е., без учета вклада электронов в их же экранировку, что ведет к хорошо известному ограничению существования ионно-звуковых колебаний (57), полученному при игнорировании роли электронов в своем собственном экранировании, т.е., при $\zeta \to 1$ полученные нами результаты полностью трансформируются в прежние.

## ЛИТЕРАТУРА